\documentclass{IEEEtran}

\usepackage[xindy]{glossaries}
\usepackage{amsfonts,mathtools,bm,amsthm}
\usepackage[OT1]{fontenc}
\usepackage[inline]{enumitem}
\usepackage{siunitx}
\usepackage{caption,subcaption}
\usepackage[hidelinks]{hyperref}
\usepackage[dvipsnames]{xcolor}
\usepackage{cleveref}
\usepackage{algorithm}
\usepackage{algpseudocode}

\newtheorem{theorem}{Theorem}
\newtheorem{proposition}{Proposition}
\newtheorem{lemma}{Lemma}

\theoremstyle{definition}

\newtheorem{example}{Example}

\theoremstyle{remark}
\newtheorem{remark}{Remark}

\def\eqns#1{\begin{equation*}#1\end{equation*}}
\def\eqnl#1#2{\begin{equation}\label{#1}#2\end{equation}}
\def\eqnsa#1{\begin{subequations}\begin{align*}#1\end{align*}\end{subequations}}
\def\eqnsml#1{\begin{multline*}#1\end{multline*}}
\def\eqnmla#1#2{\begin{subequations}\label{#1}\begin{align}#2\end{align}\end{subequations}}

\def\zero{\mathbf{0}}
\def\one{\mathbf{1}}

\def\bsF{\bm{F}}
\def\bsI{\bm{I}}

\def\boL{\mathbf{L}}

\def\bsP{\bm{\Sigma}}
\def\bsQ{\bm{Q}}

\def\boX{\mathbf{X}}
\def\boY{\mathbf{Y}}
\def\bsvphi{\bm{\varphi}}

\def\calB{\mathcal{B}}

\def\calF{\mathcal{F}}
\def\calL{\mathcal{L}}
\def\calN{\mathcal{N}}
\def\calO{\mathcal{O}}
\def\calP{\mathcal{P}}
\def\calS{\mathcal{S}}
\def\calU{\mathcal{U}}
\def\calX{\mathcal{X}}

\def\bbE{\mathbb{E}}
\def\bbN{\mathbb{N}}
\def\bbP{\mathbb{P}}
\def\bbR{\mathbb{R}}
\def\bbT{\mathbb{T}}

\def\c{\mathrm{c}}
\def\d{\mathrm{d}}

\def\s{\mathrm{s}}

\def\defeq{\doteq}
\def\ind#1{\one_{#1}}
\def\given{\,|\,}
\def\st{:}

\def\set{\leftarrow}
\def\AND{\quad\text{and}\quad}

\DeclareMathOperator*{\argmax}{argmax}

\def\oI{\bar{I}}
\def\oP{\bar{P}}
\def\oQ{\bar{Q}}
\def\oS{\bar{S}}
\def\ninf#1{\|#1\|_{\infty}}
\def\tf{\varphi}
\def\bstf{\bsvphi}

\def\best#1{\textcolor{orange}{$\mathbf{#1}$}}
\def\sbest#1{\textcolor{orange}{$#1$}}
\def\bestL#1{\textcolor{RoyalBlue}{$\mathbf{#1}$}}
\def\sbestL#1{\textcolor{RoyalBlue}{$#1$}}

\newacronym{wrt}{w.r.t.\@}{with respect to}
\newacronym{opm}{o.p.m.\@}{\emph{outer probability measure}}
\newacronym{cdf}{c.d.f.\@}{cumulative distribution function}
\newacronym{pmf}{p.m.f.\@}{probability mass function}
\newacronym{pdf}{p.d.f.\@}{probability density function}
\newacronym{rmse}{RMSE}{root mean square error}
\newacronym{map}{MAP}{maximum a posteriori}
\newacronym{smc}{SMC}{sequential Monte Carlo}
\newacronym{hmm}{HMM}{hidden Markov model}

\title{Sequential Monte Carlo algorithms for a class of outer measures}

\author
  {
    Jeremie Houssineau%
    \thanks{J.~Houssineau is with DSAP, National University of Singapore, SG. Email:~\href{mailto:stahje@nus.edu.sg}{stahje@nus.edu.sg}}
    and
    Branko Risti\'c%
    \thanks{B.~Ristic is with the School of Engineering, RMIT University, AU. Email:~\href{branko.ristic@rmit.edu.au}{branko.ristic@rmit.edu.au}}
}

\begin{document}

\maketitle

\begin{abstract}
Closed-form stochastic filtering equations can be derived in a general setting where probability distributions are replaced by some specific outer measures. In this article, we study how the principles of the sequential Monte Carlo method can be adapted for the purpose of practical implementation of these equations. In particular, we explore how sampling can be used to provide support points for the approximation of these outer measures. This step enables practical algorithms to be derived in the spirit of particle filters. The performance of the obtained algorithms is demonstrated in simulations and their versatility is illustrated through various examples.
\end{abstract}

\section*{Notations}

\begin{itemize}[leftmargin=.15\columnwidth]
\item[$\boX$:] State space
\item[$p$:] Probability distribution
\item[$f$:] Possibility functions
\item[$P$:] Probability distributions on possibility functions
\item[$\oP$:] Outer measure induced by $P$
\item[$X$:] Uncertain variable
\item[$\{W_i\}$:] Probabilistic weights ($\sum_i W_i = 1$)
\item[$\{w_i\}$:] Possibilistic weights ($\max_i w_i = 1$)
\end{itemize}

\section{Introduction}

The way uncertainty about a system of interest is modelled can greatly affect the performance of the corresponding estimation algorithms. It has been well recognised that in addition to uncertainty caused by randomness, there is another type of uncertainty,  the epistemic uncertainty, caused by the lack of knowledge \cite{OHagan2004,Benavoli2011}. The differences between epistemic and aleatoric uncertainties have been studied intensively in the field of expert systems and artificial intelligence \cite{Klir1999}, but significantly less in statistics \cite{Hampel2009}. Different methods such as Bayesian non-parametric models \cite{Ferguson1973} allow for acknowledging that all the parameters in the selected dynamical and observation processes might not be perfectly known, however, these often involve even more parameters in order to describe what is the uncertainty on the original ones, thus only offering a partial solution to the problem. Alternative modelling of uncertainties are available through the different generalisations of probability theory that have been proposed in the last 50 years, such as fuzzy logic, imprecise probabilities, possibility theory, fuzzy random sets and Dempster-Shafer theory \cite{Zadeh1965,Walley1991,Dempster1967,Shafer1976,Dubois1983,Yen1990,Friedman2001}. Most of these approaches offer the ability to model a complete absence of information but do not provide a general way of dealing with stochastic filtering.

Recently a new framework for modelling uncertainty has been introduced \cite{Houssineau2015,Houssineau2016_dataAssimilation}, which builds upon the standard measure-theoretic concept of outer measure. In particular, a specific class of outer measures based on functional integrals of the supremum has been shown to enable dynamical systems to be modelled \cite{Houssineau2017}. By combining a probability measure on a specific function space and a supremum on these functions, this class of outer measure encompasses a wide range of uncertainties, from the complete absence of knowledge to the refined information given by a probability measure on the state space. Since closed-form estimations algorithms can be derived from this framework \cite{Houssineau2017}, it is natural to inquire about the ability to implement the corresponding recursions and solve practical stochastic filtering problems without making strong assumptions on either the underlying probability measures on function spaces or on the functions themselves. This aspect of the problem has connections with robust filtering \cite{ElGhaoui2001,Geromel2002,Yin2015} in the sense that a more encompassing model should intuitively reduce the sensitivity to model discrepancies, particularly in the dynamical and observation noise.

\Gls{smc} algorithms, also referred to as particle filters, have become ubiquitous in Bayesian estimation in the last 20 years since the seminal work \cite{Gordon1993}, and have early on been used in a wide spectrum of applications \cite{Doucet2001,Cappe2007,Kantas2009,Creal2012}. Their properties have been studied in details \cite{DelMoral2004} and extensions are now countless. We will consider its simplest form in this article, known as the bootstrap particle filter, which propagates an approximation of the filtering distribution via an empirical measure of the form $N^{-1}\sum_{i=1}^N \delta_{x_i}$, where $\{x_i\}_{i=1}^N$ is a collection of samples, or particles. Since the particles are equally weighted in the case of the bootstrap particle filter, it is solely their distribution that approximately encodes the information of interest. The bootstrap particle filter yield the true filtering distribution in the limit where the number of particles tend to infinity \cite{DelMoral2004} so that new flavours of the particle filters have to be shown to have the same property.

This article proposes a practical implementation of the outer-measure recursions for stochastic filtering using an adaptation of the \gls{smc} method for propagation of support points (also referred to as particles). Although other approaches such as grid-based methods could be considered, the proposed approach is expected to inherit from the versatility and the adaptivity of the particle filter and has the potential of improving its robustness. The key problem is how to perform sampling for specific functions that represent the considered class of outer measures. Although the proposed algorithms will be very similar to a particle filter, the underlying principles will appear to differ significantly. In particular, samples will always be weighted and it is solely these weights that will carry the information. It follows that justifications for the proposed approach will take a completely different form from the ones usually used in the \gls{smc} methodology.

\section{Uncertainty and outer measures}


The objective in this section is to describe the uncertainty about a system represented by its state in a state space $\boX$, which might be a subset of $\bbR^d$ for some $d > 0$. Even when the state of the system in $\boX$ is deterministic, the standard way of representing the uncertainty about it in the Bayesian formalism is to associate a random variable $X$ with it. Formally, $X$ is assumed to be a measurable mapping from a sample space $(\Omega,\calF,\bbP)$ to the state space $\boX$ (equipped with an appropriate $\sigma$-algebra). One can think of any point $\omega \in \Omega$ as a possible state of the world in which case $X$ is simply ``extracting'' the state of the system of interest from $\omega$. Events regarding this system, say $X \in B$ for some measurable subset $B$ of $\boX$, can be expressed as events in $\Omega$ via the subset $X^{-1}(B)$ of $\calF$ and their probability can be assessed via $\bbP$. The law $p$ of $X$ is defined in this way: $p(B) = \bbP(X^{-1}(B))$ for any measurable subset $B$. In the absence of randomness, there is no need to define the probability measure $\bbP$ or the $\sigma$-algebra $\calF$, however the sample space $\Omega$ can still be considered to represent the possible states of the world.\footnote{the framework introduced here does not depend on this interpretation}

We consider the reverse construction and assume that we have been given some information about the considered system in the form of a set function $\oP$ on $\boX$ such that $\bar{P}(B) \in [0,1]$ is the \emph{credibility} of the event $X \in B$. Note that in this case, an event can be based on any subset of the considered space since there is no question of measurability and we refer to $X$ as an \emph{uncertain variable} to emphasize this. For the same reasons as in standard probability theory, we require that $\oP(\emptyset) = 0$ and that $\oP(B') \leq \oP(B)$ whenever $B' \subseteq B$. However, since we want to consider situations where there might be no objection against neither $X \in B$ nor $X \in B'$ even if $B$ and $B'$ are disjoint, we relax the assumption of additivity and instead consider that $\oP(B \cup B') \leq \oP(B) + \oP(B')$ for any subsets $B$ and $B'$. Extending this assumption to countable unions of subsets, it follows that $\oP$ is an \emph{outer measure}. Since we are interested in the case where the measure of the whole space $\boX$ is $1$, we naturally refer to this sort of outer measure as \gls{opm}.

Instead of pushing forward the probability measure $\bbP$ to $\boX$ through a random variable in order to define its law, we pull the \gls{opm} $\oP$ back onto $\Omega$ in order to define another \gls{opm} $\bar\bbP$ as
\eqns{
\bar\bbP(A) = \oP(X(A))
}
for any $A \subseteq \Omega$ (this operation would not be valid in general with probability measures). As is standard in probability theory, we consider the event $X \in B$ as a function of $\Omega$ as follows
\eqns{
(X \in B)(\omega) =
\begin{cases*}
1 & if $X(\omega) \in B$ \\
0 & otherwise,
\end{cases*}
}
which allows for writing $\bar\bbP(X \in B)$ for the credibility of the event $X \in B$. As opposed to random variables and probability measures, uncertain variables do not induce a unique \gls{opm} on the corresponding state space. This is meaningful since we might be given different pieces of information about the same uncertain quantity. We instead say that an \gls{opm} $\oP$ \emph{describes} an uncertain variable $X$ whenever $\oP(B)$ is equal to the perceived credibility of the event $X \in B$.

In some situations, it is useful to see $\oP$ as defining an upper bound for probability distributions. Indeed, it holds that $p(B) \leq \oP(B)$ for all $B$ in some $\sigma$-algebra, for some probability measure $p$ on $\boX$, in which case we will say that $p$ is \emph{(globally) bounded} by $\oP$. The credibility $\oP(B)$ can then be seen as the maximum probability for the event $X \in B$. For instance, if $\boX = \{\mathrm{head}, \mathrm{tail}\}$ then we can interpret an \gls{opm} $\oP$ defined by $\oP(\mathrm{head}) = 1$ and $\oP(\mathrm{tail}) = \alpha$ as providing the information ``the probability of head is unknown and the probability of tail is no more than $\alpha$''. If $\oP(\mathrm{head}) + \oP(\mathrm{tail}) = 1$ then $\oP$ \emph{is} a probability measure. This example illustrates the fact an \gls{opm} can characterise a unique probability distribution.

In order to solve practical problems, it is important to restrict our attention to \glspl{opm} of a specific form, and one of the simplest forms is given by
\eqns{
\oP(B) = \sup_{x \in B} f(x)
}
for any $B \subseteq \boX$, where $f$ is a positive function with supremum equal to one. This type of function is called a ``possibility distribution'' in the context of possibility theory \cite{Dubois2015}. However, since we will be using the term ``distribution'' to refer to probability distributions, we call $f$ a \emph{possibility function} instead. The set of possibility functions on $\boX$ is denoted $\boL(\boX)$. 


Although \glspl{opm} induced by possibility functions might be sufficient in some contexts, it is possible to extend significantly the sort of information that can be represented by considering \glspl{opm} of the form \cite{Houssineau2015}
\eqnl{eq:outerMeasSet}{
\oP(B) = \int \sup_{x \in B}(f) P(\d f),
}
where $P$ is a probability measure on $\boL(\boX)$. This form is suitable when little is known about the considered system; yet $\oP$ can still be as precise as a probability measure in the limit where all the possibility functions in the support of $P$ are of the form $\ind{x}$ for some $x \in \boX$. For technical reasons, we need to define the measure given by $\oP$ to any function $\tf$ in the set $\boL^{\infty}(\boX)$ of non-negative bounded functions on $\boX$ as
\eqnl{eq:outerMeas}{
\oP(\tf) = \int \ninf{\tf \cdot f} P(\d f)
}
where $\ninf{\cdot}$ is the supremum norm and where $\tf \cdot f$ denotes the point-wise product between $\tf$ and $f$, i.e.\ $(\tf \cdot f)(x) = \tf(x) f(x)$ for any $x \in \boX$. Notice that the definition in \cref{eq:outerMeasSet} can be recovered from \cref{eq:outerMeas} by considering $\tf = \ind{B}$ with $\ind{B}$ is the indicator of $B$. Whenever an \gls{opm}, e.g.\ $\oP$ or $\oP_{t|t-1}$, will be introduced, the associated probability measure $P$ or $P_{t|t-1}$ on possibility functions will be assumed to be implicitly defined.

The last ingredient in the practical definition of an \gls{opm} is the specification of one or more possibility functions. For this purpose, it is important to notice that most of the common probability density functions can be turned into possibility functions. For instance, a \emph{Gaussian possibility function} is a function $f$ in $\boL(\bbR^d)$ that verifies
\eqnl{eq:GaussianPossibility}{
f(x) = \bar\calN(x; \mu,\bsP) \defeq \exp\Big(-\frac{1}{2} (x-\mu)^T \bsP^{-1}(x-\mu)\Big),
}
for some $\mu \in \bbR^d$ and for some $d\times d$ positive-definite matrix $\bsP$ with real coefficients. Although we have not defined the notion of mean and variance for possibility functions, it is still a helpful abuse of language to refer to $\mu$ and $\bsP$ as the mean and variance of the possibility function $\bar\calN(x;\mu,\bsP)$.

\begin{remark}
If a probability distribution $P$ on $\boL(\boX)$ is supported by the indicator functions of rectangles then the proposed approach can be related to the set-membership estimation framework \cite{Jaulin2006}. In particular, connections with box-particle filtering \cite{Abdallah2008,Gning2011} can be made when $P$ is approximated by a set of samples/particles.
\end{remark}

\begin{example}
Assume that the objective is to locate a person in Melbourne. It is known that the person texted the following message: ``I'm about to see a movie on Chapel Street''. Part of the challenge is that there are three cinemas on Chapel street. Assuming the average daily number of visitors is known for each cinema, say $n_1$, $n_2$ and $n_3$, then this information can be embedded in an appropriate \gls{opm}
\eqns{
\oP(\tf) = \dfrac{n_1}{n} \ninf{ \tf \cdot \ind{C_1}} + \dfrac{n_2}{n} \ninf{ \tf \cdot \ind{C_2}} + \dfrac{n_3}{n} \ninf{ \tf \cdot \ind{C_3}}.
}
where $C_1$, $C_2$ and $C_3$ are disjoint subsets of $\boX = \bbR^2$ describing the extent of each cinema and where $n = n_1+n_2+n_3$. The associated probability measure $P$ is
\eqns{
P = \dfrac{n_1}{n} \delta_{\ind{C_1}} + \dfrac{n_2}{n} \delta_{\ind{C_2}} + \dfrac{n_3}{n} \delta_{\ind{C_3}}.
}
\end{example}

\begin{example}
Consider the information provided by a bearings-only observation produced by an idealised passive radar/sonar and taking the form of a probability distribution $p$ on the interval $\boY = (-\pi;\pi]$. This distribution can be written as a probability distribution $P$ on $\boL(\boY)$ characterised by the fact that $P$ gives mass $p(\d y)$ to the function $\ind{y}$ and gives mass $0$ to all other functions. Assume that the objective is to express the underlying uncertainty on the space $\boX = \bbR^2$ of 2-dimensional Cartesian coordinates (assumed centred on the sensor). The elements $x \in \boX$ are written as $x = (x_1,x_2)$. The corresponding outer measure $\oP$ on $\boX$ verifies
\eqns{
\oP(\tf) = \int \sup_{x \in \boX} \Big( \tf(x) \ind{y}(\arctan(x_2/x_1)) \Big) p(\d y).
}
The expression of $\oP$ can be interpreted intuitively as follows: the true state is on the half-line
\eqns{
\{x \in \boX \st \arctan(x_2/x_1) = y\}
}
with probability $p(\d y)$ for any $y \in \boY$, but there is a complete absence of knowledge on the actual position on a given half-line. This is an example of outer measure based on uncountably many possibility functions.
\end{example}

Most scenarios of interest involve several uncertain quantities and the relation between these quantities must be described, e.g.\ with joint random variables in the standard approach. Let $\boY$ be another space, let $Y$ be an uncertain variable on $\boY$ and let $\oP$ be an \gls{opm} on $\boX \times \boY$ representing the \emph{joint uncertain variable} $(X,Y) : \Omega \to \boX\times\boY$, i.e.\ $\oP(A\times B)$ is the credibility of the joint event $(X \in A, Y \in B)$. In particular, the uncertain variables $X$ and $Y$ are said to be independently described if there exist two \glspl{opm} $\oP_X$ and $\oP_Y$ such that
\eqns{
\oP(A \times B) = \oP_X(A)\oP_Y(B),
}
for any $A \subseteq \boX$ and any $B \subseteq \boY$. This property simply implies that the information we hold about $X$ and $Y$ is not interdependent. For instance, an \gls{opm} $\oP$ constructed from the information ``$X$ is $100m$ away from $Y$'' would not describe $X$ and $Y$ independently.

\section{Filtering equations and recursion}
\label{sec:filteringEquations}

Let $X_t$ be the uncertain variable on the state space $\boX_t$ describing the state at time $t \in \bbT = \{0,\dots,T\}$ with $T \in \bbN$. The observation at time $t$ is similarly modelled by an uncertain variable $Y_t$ on the observation space $\boY_t$. We consider the filtering equations
\eqnsa{
X_t & = F_t(X_{t-1},U_t) \\
Y_t & = O_t(X_t,U'_t),
}
where $F_t$ and $O_t$ are the functions describing the dynamics and the observation respectively and where $\{U_t\}_{t \in \bbT}$ and $\{U'_t\}_{t \in \bbT}$ are collections of independently described uncertain variables.

In order to describe conditional information, we first introduce a \emph{conditional possibility function} $g(\cdot \given x)$ from $\boX_{t-1}$ to $\boX_t$ describing the transition from $X_{t-1} = x$ to $X_t$ and such that $g(\cdot \given x) \in \boL(\boX_t)$ for any $x \in \boX_{t-1}$. Conditional possibility functions verify the same type of properties as conditional probability distributions: if $f_{t-1}$ is a possibility function on $\boX_{t-1}$ describing $X_{t-1}$ then
\eqnl{eq:singlePosPrediction}{
f_t(x) = \sup_{x' \in \boX_{t-1}} g(x \given x') f_{t-1}(x')
}
is a possibility function describing $X_t$. This prediction equation for possibility functions is the analogue of the Chapman-Kolmogorov equation in standard Bayesian filtering except that the integral is replaced by a supremum and probability density functions are replaced by possibility functions.

\begin{remark}
Although possibility functions can be seen as renormalised probability distributions, this identification cease to hold when applying operations such as \cref{eq:singlePosPrediction}. For instance, if $\boX = \{-2,-1,0,1,2\}$, if $f_{t-1} = \ind{\{-1,1\}}$ and if
\eqns{
g(x \given x') = 
\begin{cases*}
1 & if $x = x'$ \\
1/2 & if $|x-x'| = 1$ \\
0 & otherwise,
\end{cases*}
}
then $f_t(x)$ is equal to $1$ for $x = -1, 1$ and to $1/2$ for $x = -2, 0, 2$. This is different from the result that would be obtained if the standard Chapman-Kolmogorov equation was applied to normalised version of $f_{t-1}$ and $g$ and if the result was turned back into a possibility function (indeed we would have $f_t(0) = 1$).
\end{remark}

Let $\oQ_t(\cdot \given X_{t-1} = x)$ be a conditional \gls{opm} on the state space $\boX_t$ representing the uncertainty induced by $F_t(x, U_t)$, e.g.\ $\oQ_t(\ind{B} \given X_{t-1} = x)$ is the credibility of the event $X_t \in B$ given that $X_{t-1} = x$. Following \cite{Houssineau2018_detection}, we consider that $\oQ_t(\cdot \given X_{t-1} = x)$ is of the form
\eqns{
\oQ_t(\tf \given X_{t-1} = x) = \int \ninf{\tf \cdot g(\cdot \given x)} Q_t(\d g \given X_{t-1}) 
}
where $Q_t(\cdot \given X_{t-1})$ is a probability measure on conditional possibility functions which does not depend on the realisation $x$ of $X_{t-1}$ (the conditioning is only indicated in order to underline the nature of the possibility functions in the support of $Q_t(\cdot \given X_{t-1})$).

Similarly, we denote by $\oS_t(\cdot \given X_t=x)$ the conditional \gls{opm} on $\boY_t$ describing the uncertainty induced by $O_t(x,U'_t)$. The \gls{opm} $\oS_t$ describes the knowledge about the point observation in $\boY_t$ given the state in $\boX_t$. The relation between $\oS_t(\cdot \given X_t = x)$ and the likelihood will be detailed later in this section. 

In general, we might not directly receive the realisation $y_t$ of the observation variable $Y_t$ at time $t$. Information about~$y_t$ might be given instead under a different form, e.g.\ as a natural language statement or as an event such as $y_t \in A$. The latter case can model information provided by digital sensors when a given pixel or resolution cell is known to contain the point observation $y_t$. The information about the point observation~$y_t$ is referred to as \emph{observed information} and is represented by an \gls{opm} $\oI_t$ on $\boY_t$.

\begin{remark}
Although it is unusual to assume that a point observation $y_t$ is not directly received, this is a convenient approach when dealing with non-standard observations. For instance, in the case of natural language statements, it is not easy to formally define the space of all possible statements so instead we assume that $y_t$ is the point observation corresponding to what is perceived by the person emitting the statement and the statement itself is simply considered as information about $y_t$.
\end{remark}

The following theorem describes the prediction from time $t-1$ to time $t$, where $\oP_{t-1|t-1}$ denotes the posterior \gls{opm} at the previous time step, i.e.\ the \gls{opm} describing $X_{t-1}$ given the observed information $\oI_0, \dots, \oI_{t-1}$, and where $P_{t-1|t-1}$ is the probability measure on $\boL(\boX_t)$ associated with the \gls{opm} $\oP_{t-1|t-1}$. 

\begin{theorem}
\label{thm:predition}
The predicted \gls{opm} $\oP_{t|t-1}$, which describes the uncertain variable $X_t$ given the observed information $\oI_0,\dots,\oI_{t-1}$, is characterised by
\eqns{
\oP_{t|t-1}(\tf) = \int \ninf{\tf \cdot \zeta_t(f,g)} Q_t(\d g \given X_{t-1}) P_{t-1|t-1}(\d f)
}
for any $\tf \in \boL^{\infty}(\boX_t)$, where the possibility function $\zeta_t(f,g)$ on $\boX_t$ is defined as
\eqns{
\zeta_t(f,g)(x) = \sup_{x' \in \boX_{t-1}} g(x \given x') f(x').
}
for any $x \in \boX_t$.
\end{theorem}

The proof of \cref{thm:predition} can be found in the appendix, together with the proofs of the other results in the article.

The mapping $\zeta_t$ defined in \cref{thm:predition} takes a possibility function $f$ in $\boL(\boX_{t-1})$ and incorporates the uncertainty brought by the conditional possibility function $g$ into it so that $\zeta_t(f,g) \in \boL(\boX_t)$ represents the resulting uncertainty at time $t$. 
The probability distribution $P_{t|t-1}$ on which the \gls{opm} $\oP_{t|t-1}$ is based gives probability mass $Q_t(\d g \given X_{t-1}) P_{t-1|t-1}(\d f)$ to the function $\zeta_t(f,g)$.

The update mechanism is derived in the next theorem for the considered setting where the uncertainty induced by $U'_t$ is not assumed negligible, as opposed to \cite{Houssineau2017}. As before, the probability measures $S_t(\cdot \given X_t)$ and $I_t$ are the ones underlying the \glspl{opm} $\oS_t(\cdot \given X_t = x)$ and $\oI_t$.

\begin{theorem}
\label{thm:generalUpdate}
The posterior \gls{opm} $\oP_{t|t}$ on $\boX_t$ resulting from the update of the predicted \gls{opm} $\oP_{t|t-1}$ on $\boX_t$ by the observed information $\oI_t$ on $\boY_t$ can be expressed as
\eqns{
\oP_{t|t}(\tf) = \dfrac{\int \ninf{ \tf \cdot f \cdot \zeta'_t(s,h)} P_{t|t-1}(\d f)S_t(\d s \given X_t) I_t(\d h)}{\int \ninf{f \cdot \zeta'_t(s,h)} P_{t|t-1}(\d f)S_t(\d s \given X_t) I_t(\d h)},
}
for any $\tf \in \boL^{\infty}(\boX_t)$, where the function $\zeta'_t(s,h) \in \boL^{\infty}(\boX_t)$ is characterised by
\eqnl{eq:thm:generalUpdate}{
\zeta'_t(s,h)(x) = \ninf{ s(\cdot \given x) \cdot h } 
}
for all $x \in \boX_t$.
\end{theorem}


The result of \cref{thm:generalUpdate} can be simplified by considering the case where all the involved \glspl{opm} are based on a single-possibility function, i.e.\ when there exist $f_t \in \boL(\boX_t)$, $h_t \in \boL(\boY_t)$ and a conditional possibility function $s_t( \cdot \given x)$ on $\boY_t$ such that
\eqnsa{
\oP_{t|t-1}(\tf) & = \ninf{\tf \cdot f_{t|t-1}}, \\
\oI_t(\tf) & = \ninf{ \tf \cdot h_t}, \\
\oS_t(\tf \given X_{t-1} = x) & = \ninf{\tf \cdot s_t(\cdot \given x)},
}
for any $\tf \in \boL^{\infty}(\boX_t)$ and any $x \in \boX_t$. It follows in this simplified setting that the posterior \gls{opm} at time $t$ verifies $\oP_{t|t}(\tf) = \ninf{\tf \cdot f_{t|t}}$ for some possibility function $f_{t|t}$ on $\boX_t$, defined as
\eqns{
f_{t|t}(x) = \dfrac{ f_{t|t-1}(x) \ninf{ h_t \cdot s_t(\cdot \given x) } }{ \sup_{x' \in \boX_t} f_{t|t-1}(x') \ninf{ h_t \cdot s_t(\cdot \given x') } }.
}
In particular, if a point observation $y_t$ is made available, then $h_t = \ind{y_t}$ and
\eqnl{eq:singlePosUpdate}{
f_{t|t}(x) = \dfrac{ f_{t|t-1}(x) s_t(y_t \given x)  }{ \sup_{x' \in \boX_t} f_{t|t-1}(x') s_t(y_t \given x') },
}
which is the analogue of Bayes' theorem with a supremum instead of an integral and with possibility functions rather than probability density functions. 

\begin{remark}
Following the same approach as in \cite[Theorem~10]{Houssineau2017}, it can be proved that the filtering equations \cref{eq:singlePosPrediction} and \cref{eq:singlePosUpdate} lead to the same recursion as the Kalman filter in terms of mean and variance when all the involved possibility functions are Gaussian.
\end{remark}

It is assumed in the rest of the paper that a point observation $y_t \in \boY_t$ is made available so that the observed information takes the form $\oI_t(\tf) = \ninf{\tf \cdot \ind{y_t}} = \tf(y_t)$, which corresponds to $I_t = \delta_{\ind{y_t}}$. It is also assumed that the conditional \gls{opm} $\oQ_t(\cdot \given X_{t-1} = x)$ is based on a single possibility function $g_t(\cdot \given x)$. These assumptions can be easily lifted and are only made for the sake of simplicity. To sum up, the filtering equations are expressed as
\eqnmla{eq:filteringUncertainTransition}{
\oP_{t|t-1}(\tf) & = \int \ninf{\tf \cdot \zeta_t(f,g_t)} P_{t-1|t-1}(\d f) \\
\oP_{t|t}(\tf) & = \dfrac{\int \ninf{ \tf \cdot f \cdot s(y_t \given \cdot)} P_{t|t-1}(\d f)S_t(\d s \given X_t) I_t(\d h)}{\int \ninf{f \cdot s(y_t \given \cdot)} P_{t|t-1}(\d f)S_t(\d s \given X_t) I_t(\d h)}
}
for any $\tf \in \boL^{\infty}(\boX_t)$.

\section[Approximating o.p.m.s]{Approximating \glspl{opm}}
\label{sec:sampling}

The two building blocks of the considered class of \glspl{opm} are probability measures and possibility functions. Approximating the former is the topic of a vast body of literature, however, it is less clear how to proceed with the latter. In the following sections, we consider separately the cases of continuous and discrete spaces.

\subsection{For possibility functions on a continuous space}

The general objective in this section is to devise an approximation for possibility functions that makes the above-described filtering equations tractable. One of the first solution that comes to mind is a grid-based approximation: if $f$ is a possibility function on $\boX$ and if $G$ is a partition of $\boX$ then we can approximate $f$ by a piece-wise constant function $\tilde{f}$ defined for any $A \in G$ and any $x \in A$ by
\eqns{
\tilde{f}(x) = \sup_{x' \in A} f(x').
}
This approach has been considered in \cite{Bishop2018} for inference from natural language statements. However, the usual disadvantages of grid-based approaches apply equally to such an approximated possibility function: there might be little prior knowledge about the support of $f$ so that a large area has to be covered, and this can make this method highly inefficient. One of the usual alternatives to grid-based methods in the context of Bayesian inference is the particle-based approach which relies on sampling from the probability distributions of interest. However, sampling does not apply directly to a possibility function $f \in \boL(\boX)$. Yet, this can be seen as an advantage, since we can select the probability distribution of our choice to sample from.

Putting aside the question of which probability distribution to sample from, consider that we have computed $N$ samples $\{x_i\}_{i=1}^N$ from a distribution $p$ on $\boX$. These samples can be used as support points for an approximation of $\ninf{ \tf \cdot f }$ for any $\tf \in \boL^{\infty}(\boX)$ as
\eqnl{eq:approx_by_samples}{
\ninf{ \tf \cdot f } \approx \max_{1 \leq i \leq N} w_i \tf(x_i)
}
with $w_i \propto f(x_i)$ for any $i \in \{1,\dots,N\}$ and $\max_i w_i = 1$. This approach is different in nature from the approximation of a probability distribution by the empirical measure $\sum_i \delta_{x_i}$.

\begin{remark}
One of the main objectives in practice when dealing with stochastic filtering is to find an approximation of the mean or mode of the filtering distributions, and the analogue of the latter for a possibility function $f$ is $\argmax_x f(x)$. Therefore, the interest will often be in the dual problem of locating $\argmax_x \tf(x)f(x)$ rather than approximating $\ninf{ \tf \cdot f }$.
\end{remark}

The following proposition ensures, under conditions, that the error in the approximation \cref{eq:approx_by_samples} converges to $0$ when the number of samples tends to infinity.

\begin{proposition}
\label{res:convergence}
Let $\tf \in \boL^{\infty}(\boX)$, let $f \in \boL(\boX)$, let $p$ be a probability measure on $\boX$ with the same support as $f$ and let $x_i \sim p$ for $i \in \{1,\dots,N\}$ and for some integer $N$. If the function $\tf \cdot f$ is Lipschitz and achieves its supremum then the following convergence in probability holds
\eqns{
\max_{1 \leq i \leq N} w_i \tf(x_i) \xrightarrow{N \to \infty}{} \ninf{ \tf \cdot f },
}
where $w_i = f(x_i)$ for any $i \in \{1,\dots,N\}$.
\end{proposition}

Although \cref{res:convergence} is restricted to sufficiently regular functions $\tf$ and $f$, this result illustrates the freedom on the choice of the probability distribution $p$ that we sample from. The speed at which this convergence will take place will however greatly depend on the choice of $p$. If the approximation $\ninf{\tf \cdot f}$ was performed for a fixed $\tf$ then it would be meaningful to make $p$ depend on this function, however, in the context of filtering, $\tf$ might be for instance the likelihood of future observations, which is not available when approximating the prior in general. Keeping in mind the case where $\tf$ is the likelihood of future observations, it also appears that $\tf \cdot f$ might achieve its maximum in an area where $f$ takes low values so that the samples should be sufficiently spread across the support of $f$ with less samples where $f$ is small.

If $f$ is integrable, the simplest way to define the \gls{pdf} $p$ from which to sample from is to renormalise $f$ as
\eqns{
p(x) = \dfrac{f(x)}{\int f(x) \d x}.
}
In the performance assessment in \cref{sec:results}, the \gls{pdf} $p$ will be referred to as the \emph{scaled} distribution associated with $f$. However, it is also possible to define a probability distribution~$p^*$ providing the maximum diversity of samples while being small/negligible where $f$ is. In theory, one can select $p^*$ as the solution of a constrained optimisation problem:
\eqnl{eq:maxEntropy}{
p^* = \argmax_p H(p)
}
subject to
\begin{enumerate}
\itemsep0.5em 
\item $p$ is a probability distribution on $\boX$
\item \label{eq:entropyConditionOuterMeas} $\int \ind{B}(x) p(x) \d x \leq \sup_{x \in B} f(x) \text{ for any } B \in \calB(\boX)$
\end{enumerate}
where $H(p) = \bbE[-\ln p(X)]$ is the differential entropy of a probability density function $p$, with $\bbE$ the expectation w.r.t.\ a random variable $X$ with distribution $p$.

The choice of the probability distribution $p^*$ follows from the \emph{principle of maximum entropy}, first established in the context of statistical mechanics in \cite{Jaynes1957}. This principle states that the probability distribution which best represents the available information is the one with maximum entropy. Interpreting the possibility function as the available information, the formulation \cref{eq:maxEntropy} follows directly. The principle of maximum entropy can be used within the Bayesian framework to determine prior probability distributions \cite{Jaynes1968} and is therefore compatible with the proposed approach. In particular, the normal distribution $\calN(\cdot ; \mu,\sigma^2)$ is the maximum-entropy distribution with mean $\mu$ and variance $\sigma^2$ that is supported by the real line \cite{Dowson1973}; this fact supporting the common choice of a normal distribution as a prior. 

We solve the problem of \cref{eq:maxEntropy} for the Gaussian possibility function $\bar\calN(\cdot; 0,1)$ on $\bbR$.


\begin{lemma}
\label{res:monotonePossibility}
Let $f$ be a monotonically increasing possibility function defined on an interval $I = (-\infty,b]$ for some $b \in \bbR$, then a probability distribution $p$ on $I$ is bounded by $f$ if and only if its \gls{cdf} $F$ verifies $F \leq f$.
\end{lemma}

\begin{proposition}
\label{res:maxEntropy}
The solution of \cref{eq:maxEntropy} when $f = \bar\calN(\cdot; 0,1)$ is the symmetrical probability distribution $p^*$ characterised on $(-\infty,0]$ by
\eqns{
p^*(x) =
\begin{cases*}
-\frac{1}{2}x\bar\calN(x; 0,1) & if $x < x^*$ \\
\frac{1}{2x^*}\big(\bar\calN(x^*; 0,1) - 1\big) & if $x^* \leq x \leq 0$ 
\end{cases*}
}
where $x^*$ is the strictly negative solution of
\eqnl{eq:xStar}{
\exp\Big(-\frac{1}{2}x^2\Big)(x^2 + 1) = 1.
}
\end{proposition}

\begin{figure}
\centering
\includegraphics[trim=50pt 300pt 55pt 300pt,clip,width=\columnwidth]{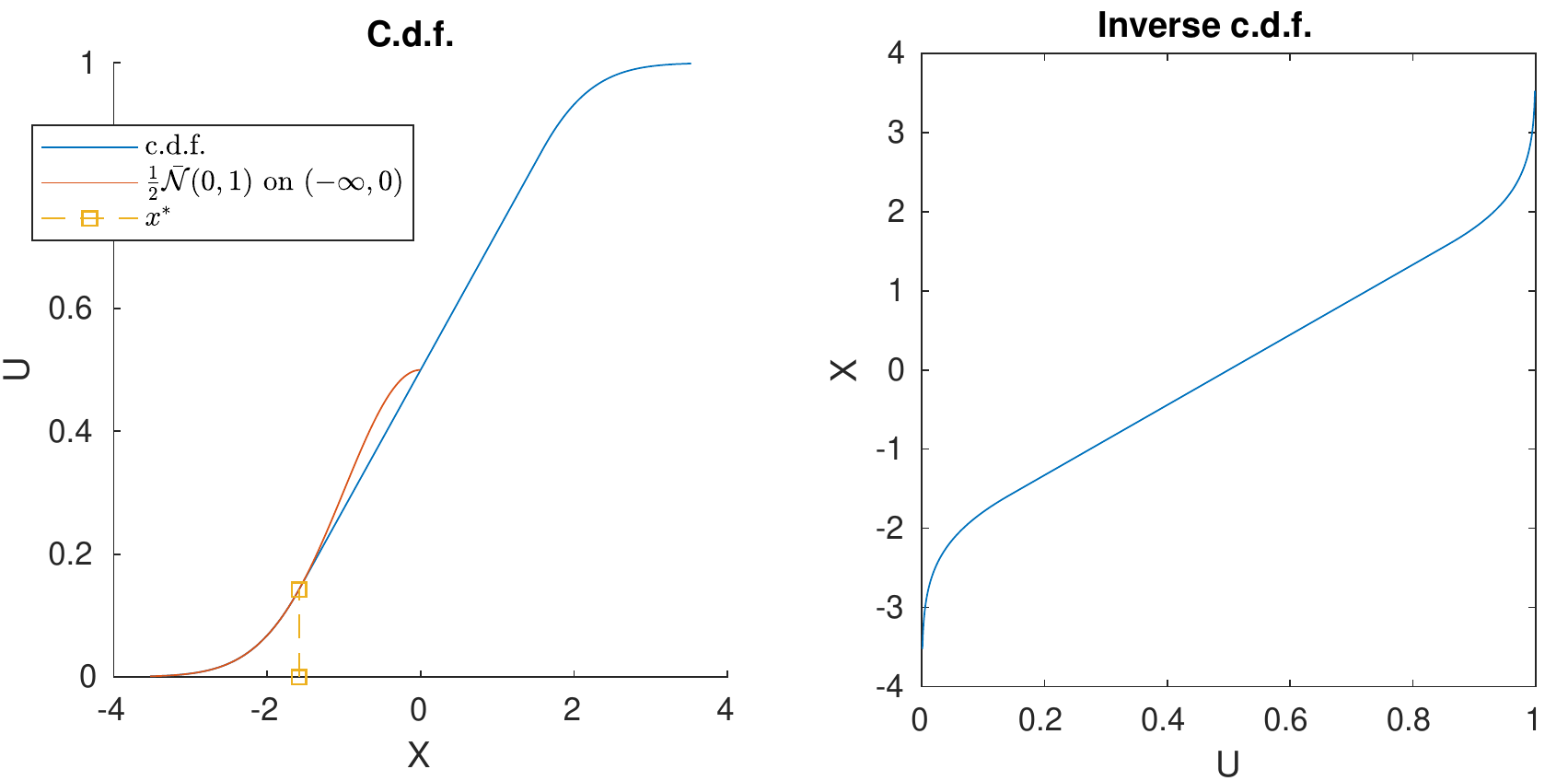}
\caption{C.d.f.\ and inverse c.d.f.\ corresponding to the solution of \cref{eq:maxEntropy} in the case of the possibility function $\bar\calN(\cdot; 0,1)$. The red line indicates the values of $\frac{1}{2}\bar\calN(\cdot; 0,1)$ on $(-\infty,0)$, which limits the distribution of half of the probability mass (by symmetry).}
\label{fig:cdfOfPStar}
\end{figure}

Notice that the solution of \cref{eq:xStar} can be easily found numerically, e.g.\ by the bisection method. The solution given in \cref{res:maxEntropy} is illustrated in \cref{fig:cdfOfPStar}. A direct consequence of this \lcnamecref{res:maxEntropy} is that random variables distributed according to $p^*$ can be easily obtained through inverse transform sampling, i.e.\ as
\eqns{
(F^*)^{-1}(U) = 
\begin{dcases*}
-\sqrt{-2\ln(2 U)} & if $u < \frac{1}{2} \bar\calN(x^*; 0,1)$ \\
\sqrt{-2\ln(2 (1-U))} & if $u > 1 - \frac{1}{2} \bar\calN(x^*; 0,1)$ \\
\dfrac{(2U-1)x^*}{1 - \bar\calN(x^*; 0,1)} & otherwise.
\end{dcases*}
}
where the law of $U$ is the uniform distribution on $[0,1]$, denoted $\calU([0,1])$, and where $F^*$ is the \gls{cdf} of $p^*$. Henceforth, this method will be referred to as the \emph{global entropy} method.

In order to keep general the algorithm description given in the next sections, we will denote by $\calP_{\c}(f)$ the probability distribution from which samples are obtained for the approximation of a given possibility function $f$ on a continuous space, regardless of the method used (i.e.\ scaled or global entropy). The different methods for defining $\calP_{\c}(f)$ will be evaluated in \cref{sec:results}.

\subsection{For possibility functions on a discrete space}

Possibility functions on discrete spaces might not need to be directly approximated since a large number of their values can be simply stored without inducing computational issues. However, when dealing with possibility functions of the form of a \emph{max-mixture}, e.g.\
\eqnl{eq:maxMixture}{
f(x) = \max_{1 \leq i \leq N} w_i f_i(x)
}
for some collections $\{w_i\}_{i=1}^N$ and $\{f_i\}_{i=1}^N$ of scalars in $[0,1]$ and possibility functions on $\boX$ respectively, it is convenient to first select one component $i \in \{1,\dots,N\}$ at random and then approximate the corresponding possibility function $f_i$ as previously. It is clearly possible to simply renormalise the possibility function on $\{1,\dots,N\}$ associated with $\{w_i\}_{i=1}^N$, and define the associated \gls{pmf} via
\eqns{
W_i = \dfrac{w_i}{\sum_{j=1}^N w_j}.
}
As in the continuous case, this will be referred to as the \emph{scaled} distribution. The global entropy approach used to deal with continuous spaces can also be applied to discrete spaces. In order to compute the corresponding \gls{pmf}, it is sufficient to sort the points in the collection $\{w_i\}_i$ in increasing order and, assuming for the sake of simplicity that the $w_i$'s are already sorted, to calculate the associated \gls{pmf} in the following way:
\eqnl{eq:discr_global_entropy}{
W_i = \max_{i \leq j \leq N} \dfrac{w_j - \sum_{k=1}^{i-1} W_k}{j-i+1},
}
for any $i \in \{1,\dots,N\}$, where $\sum_{k=1}^0 W_k = 0$ by convention. This is simply the maximum mass one can attribute to the point $i$ while leaving enough probability mass for the next points. \Cref{eq:discr_global_entropy} is the global entropy method for discrete spaces.

\begin{figure}
\centering
\includegraphics[width=.75\columnwidth]{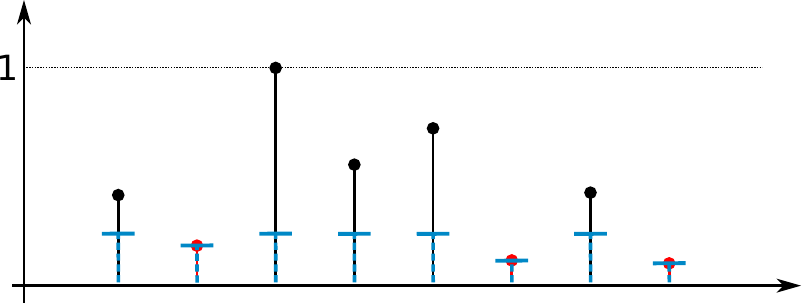}
\caption{Water-pouring operation for a discrete distribution (the blue weights sum to one). Red markers are used when the possibility function and the induced \gls{pmf} are equal.}
\label{fig:waterPouringDiscrete}
\end{figure}

Although $\{W_i\}_{i=1}^N$ as defined in \cref{eq:discr_global_entropy} indeed  corresponds to the \gls{pmf} with maximum entropy that is bounded by the $w_i$'s, i.e.\
\eqns{
\sum_{i \in B} W_i \leq \sup_{i \in B} w_i
}
for any subset $B \subseteq \{1,\dots,N\}$, it is possible to further increase the entropy by only requiring that $W_i \leq w_i$ for any $i \in \{1,\dots,N\}$, in which case we say that the associated \gls{pmf} is \emph{locally bounded}\footnote{Local boundedness is a weaker constraint than global boundedness so that the former allows for a larger entropy than the latter.} by $w$. This technique will be referred to as the \emph{local entropy} method in the following sections. Finding the \gls{pmf} with maximum entropy that is locally bounded by the $w_i$'s can be seen as a \emph{water pouring} operation as illustrated on \cref{fig:waterPouringDiscrete}. This operation is easier to justify for discrete probability distributions than for continuous ones since it is not always applicable in the latter case, e.g.\ when $\int f(x) \d x < 1$. Although this local approach is ad-hoc, the objective is simply to preserve as many terms in \cref{eq:maxMixture} as possible, which makes attractive the local viewpoint.

The probability distribution used to obtain support points for the approximation of the possibility function associated with the collection of weights $\{w_i\}_{i=1}^N$ is denoted $\calP_{\d}\big(\{w_i\}_{i=1}^N\big)$. The different options for defining such a probability distribution on a discrete space will be evaluated in \cref{sec:results}.

\subsection[For o.p.m.s]{For \glspl{opm}}

The objective is now to detail a procedure yielding a $N$-sample approximation $\oP^{\s}$ of a given \gls{opm} $\oP$ on a given space~$\boX$, where ``$\s$'' stands for ``sampled''. This approximation can be expressed for any $\tf \in \boL^{\infty}(\boX)$ under the form
\eqnl{eq:sampleRepresantation}{
\oP(\tf) \approx \oP^{\s}(\tf) = \sum_{i=1}^m W_i \max_{1 \leq j \leq M_i} w_{i,j} \tf(x_{i,j}),
}
where $\{W_i\}_{i=1}^m$ is a collection of positive weights summing to $1$ and where $\calX_i = \{ (w_{i,j},x_{i,j}) \}_{j=1}^{M_i}$ is a collection of positively-weighted samples for any $i \in \{1,\dots,m\}$, for some integer $m$. The collection of weights $\{W_i\}_i$ comes from the (standard) sampling of possibility functions from the probability measure $P$ on $\boL(\boX)$ that is associated with $\oP$, while $\calX_i$ comes from the proposed approximation of the sampled possibility functions. The approximation in \cref{eq:sampleRepresantation} is consistent with the functional integrals of the supremum considered before, as it displays the finite versions of the integral and the supremum, i.e.\ a sum and a maximum respectively.

\begin{algorithm}
\caption{Approximation algorithm}
\label{alg:sampling}
\begin{algorithmic}[1]
\Function{$\calL = \text{\textnormal{Approximation}}(\oP,N)$}{} 
\State $\calF=\emptyset$
\Comment{Set of approximated functions}
\State $m=0$
\Comment{Number of approximated functions}
\Repeat \label{alg:sampling:fsimP} $f \sim P$
\If{$f \notin \calF$}
\State $m \set m+1$
\State $f_m = f$
\State $M_m=1$
\State $\calF = \calF \cup \{f_m\}$
\State $x_{m,1}\sim \calP_{\c}(f_m)$
\State $\tilde{w}_{m,1} = f_m(x_{m,1})$
\State $\tilde\calX_m = \{(\tilde{w}_{m,1}, x_{m,1})\}$
\Else
\State Find $n$ s.t. $f_n\in\calF$ and $f_n=f$
\State $M_n \set M_n + 1$
\State $x_{n,M_n}\sim \calP_{\c}(f)$
\State $\tilde{w}_{n,M_n} = f(x_{n,M_n})$
\State $\tilde\calX_n = \tilde\calX_n \cup \{(\tilde{w}_{n,M_n}, x_{n,M_n})\}$
\EndIf
\Until $\sum_{\ell=1}^m M_\ell= N$
\For {$i=1,\dots,m$}
\vspace{.5em}
\State \label{alg:sampling:W} $W_i = \dfrac{\displaystyle M_i \max_{1\leq j \leq M_i} \tilde{w}_{i,j}}{ \displaystyle \sum_{i'} M_{i'} \max_{1\leq j \leq M_{i'}} \tilde{w}_{i',j}}$
\For {$j=1,\dots,M_i$}
\vspace{.5em}
\State $w_{i,j} =  \dfrac{\tilde{w}_{i,j}}{\displaystyle\max_{1\leq \ell \leq M_i} \tilde{w}_{i,\ell}}$
\EndFor
\EndFor
\State {\bf Output:} $\calL = \left\{\left(W_i,\calX_i = \{(w_{i,j},x_{i,j})\}_{j = 1}^{M_i}\right)\right\}_{i=1}^{m}$
\EndFunction
\end{algorithmic}
\end{algorithm}

Note that the approximated \gls{opm} $\oP^{\s}$ is characterised by the collection $\calL = \{ ( W_i, \calX_i) \}_{i=1}^m$. A pseudo-code of this procedure is given in \cref{alg:sampling} where the approximation of an arbitrary \gls{opm} $\oP$ is obtained iteratively by selecting a possibility function $f$ from the associated distribution $P$ on $\boL(\boX)$ and by approximating this possibility function via samples from $\calP_{\c}(f)$.

As opposed to the single-function case $\oP(\tf) = \ninf{\tf \cdot f}$ where it is simply assumed that the sample weights have maximum~$1$, the normalisation of the sample weights has an effect on the possibility-function weight $W_i$ in general (as described on line~\ref{alg:sampling:W} of \cref{alg:sampling}). This is due to the fact that $\oP^{\s}$ can only be renormalised as a whole and rescaling within its expression have to be compensated for.

\section[SMC with o.p.m.s]{\Acrlong{smc} with \glspl{opm}}
\label{sec:SMC}

Whichever approach is used for obtaining support points for the approximation of possibility functions, the overall mechanisms of propagating these weighted samples as an approximation of the sequence of posterior \glspl{opm} remain the same, and are detailed in this section.

\subsection{Initialisation}

It is assumed that the initial \gls{opm} $\oP_{0|0}$ is replaced by a $N$-sample approximation $\oP^{\s}_{0|0}$. This approximation can be expressed as
\eqns{
\oP_{0|0}(\tf) \approx \oP^{\s}_{0|0}(\tf) = \sum_{i=1}^{m_0} W^i_0 \max_{1 \leq j \leq M^i_0} w^{i,j}_0 \tf(x^{i,j}_0),
}
for some collection $\calL_0 = \{(W^i_0,\calX^i_0)\}_{i=1}^{m_0}$ with, for any $i \in \{1,\dots,m_0\}$,
\eqns{
\calX^i_0 = \big\{ \big( w^{i,j}_0,x^{i,j}_0 \big) \big\}_{j = 1}^{M^i_0}.
}
By construction, it holds that $\sum_i M^i_0 = N$.

\subsection{Prediction}
\label{ssec:prediction}

Assuming that the posterior \gls{opm} $\oP_{t-1|t-1}$ at the previous time step is approximated by
\eqns{
\oP^{\s}_{t-1|t-1}(\tf) \defeq \sum_{i=1}^{m_{t-1}} W^i_{t-1} \max_{1 \leq j \leq M^i_{t-1}} w^{i,j}_{t-1} \tf(x^{i,j}_{t-1}),
}
it follows that the predicted \gls{opm} based on $\oP^{\s}_{t-1|t-1}$ takes the form
\eqns{
\tf \mapsto \sum_{i=1}^{m_{t-1}} W^i_{t-1} \max_{1 \leq j \leq M^i_{t-1}} w^{i,j}_{t-1} \ninf{ \tf \cdot g_t(\cdot \given x^{i,j}_{t-1}) }.
}
This \gls{opm} is a sum of max-mixtures of the form \cref{eq:maxMixture} and needs to be further approximated. This is achieved by drawing a sample $\tilde{x}^{i,j}_t$ from $\calP_{\c}(g_t(\cdot \given x^{i,j}_{t-1}))$ for any index $j \in \{1,\dots,M^i_{t-1}\}$ and any $i \in \{1,\dots,m_{t-1}\}$, so that the predicted \gls{opm} $\oP_{t|t-1}$ can be approximated by
\eqns{
\oP^{\s}_{t|t-1}(\tf) = \sum_{i=1}^{m_{t-1}} W^i_{t|t-1} \max_{1 \leq j \leq M^i_{t-1}} w^{i,j}_{t|t-1} \tf(\tilde{x}^{i,j}_t),
}
where
\eqnl{eq:predW}{
W^i_{t|t-1} = \dfrac{\displaystyle W^i_{t-1} \max_{1 \leq j,l \leq M^i_{t-1}} w^{i,j}_{t-1} g_t(\tilde{x}^{i,l}_t \given x^{i,j}_{t-1})}{\displaystyle \sum_k W^k_{t-1} \max_{1 \leq j,l \leq M^k_{t-1}} w^{k,j}_{t-1} g_t(\tilde{x}^{k,l}_t \given x^{k,j}_{t-1})}
}
and
\eqnl{eq:predw}{
w^{i,j}_{t|t-1} = \dfrac{\displaystyle \max_{1 \leq k \leq M^i_{t-1}} w^{i,k}_{t-1} g_t(\tilde{x}^{i,j}_t\given x^{i,k}_{t-1})}{\displaystyle \max_{1 \leq k,l \leq M^i_{t-1}} w^{i,k}_{t-1} g_t(\tilde{x}^{i,l}_t \given x^{i,k}_{t-1})}.
}
The collections of samples $\{ \tilde{x}^{i,j}_t \}_{j=1}^{M^i_{t-1}}$ bear the subscript ``$t$'' rather than ``$t|t-1$'' since they will not be affected by the update, as will become clear in the next section. The complexity of this prediction step is quadratic in the number of samples. This is not surprising since determining the \gls{map} in a particle filter is also of quadratic complexity.

\begin{remark}
Is some situations, and in particular when the uncertainty on the motion model is small, the predicted \gls{opm} can be further approximated by considering $g_t(\tilde{x}^{i,l}_t \given x^{i,k}_{t-1}) \approx 0$ for any $i \in \{1,\dots,m_{t-1}\}$ whenever $l \neq k$. This approximation lowers the complexity of the prediction step to linear in the number of samples. This linear implementation will be referred as the $\calO(N)$ version of the proposed method in \cref{sec:results}, while the prediction given by \cref{eq:predW,eq:predw} will be called the $\calO(N^2)$ version.
\end{remark}

\subsection{Update}

The mechanisms in the update also differ from standard particle filtering. In the general case where the integral w.r.t.\ $S_t(\cdot \given X_t)$ cannot be evaluated, one has to sample $L_t$ conditional possibility functions $\{s_{\ell}\}_{\ell=1}^{L_t}$ from $S_t(\cdot \given X_t)$ in order to enable an approximation of the updated \gls{opm} $\oP_{t|t}$ to be computed. It follows that
\eqnl{eq:upPs}{
\oP_{t|t}(\tf) \approx \oP^{\s}_{t|t}(\tf) = \sum_{\ell=1}^{L_t} \sum_{i=1}^{m_{t-1}} \widetilde{W}^{i,\ell}_t \max_{1 \leq j \leq M^i_{t-1}} \tilde{w}^{i,j,\ell}_t \tf\big(\tilde{x}^{i,j}_t\big),
}
with
\eqnl{eq:upW}{
\widetilde{W}^{i,\ell}_t = \dfrac{\displaystyle W^i_{t|t-1} \max_{1 \leq j \leq M^i_{t-1}} w^{i,j}_{t|t-1} s_{\ell}(y_t \given \tilde{x}^{i,j}_t)}{\displaystyle \sum_n \sum_k W^k_{t|t-1} \max_{1 \leq j \leq M^k_{t-1}} w^{k,j}_{t|t-1} s_n(y_t \given \tilde{x}^{k,j}_t)}
}
and
\eqnl{eq:upw}{
\tilde{w}^{i,j,\ell}_t = \dfrac{w^{i,j}_{t|t-1} s_{\ell}(y_t \given  \tilde{x}^{i,j}_t)}{\displaystyle \max_{1 \leq k \leq M^i_{t-1}} w^{i,k}_{t|t-1} s_{\ell}(y_t \given \tilde{x}^{i,k}_t)},
}
which brings the total number of samples from $N$ to $N \times L_t$ and the total number of approximated possibility functions from $m_{t-1}$ to $m_t \defeq m_{t-1} \times L_t$.
The number of particles is subsequently reduced to N by application of resampling (\cref{ssec:resampling}).

If $S_t(\cdot \given X_t)$ is supported by a finite family $\{s_t^{\ell}\}_{\ell=1}^{L_t}$ of possibility functions then it holds that
\eqns{
S_t(\cdot \given X_t) = \sum_{\ell=1}^{L_t} V^{\ell}_t \delta_{s^{\ell}_t}
}
for some collection $\{V^{\ell}_t\}_{\ell=1}^{L_t}$ of weights. The sampling procedure on $S_t(\cdot \given X_t)$ can thus be avoided and only the expression of the weights $\widetilde{W}^{i,\ell}_t$ has to be changed to
\eqns{
\widetilde{W}^{i,\ell}_t = \dfrac{\displaystyle W^i_{t|t-1} V^{\ell}_t \max_{1 \leq j \leq M^i_{t-1}} w^{i,j}_{t|t-1} s_t^{\ell}(y_t \given \tilde{x}^{i,j}_t)}{\displaystyle \sum_n \sum_k W^k_{t|t-1} V^n_t \max_{1 \leq j \leq M^k_{t-1}} w^{k,j}_{t|t-1} s_t^n(y_t \given \tilde{x}^{k,j}_t)}.
}

\subsection{Resampling}
\label{ssec:resampling}

In order to focus the computational power on the areas of the state space with non-negligible likelihood, the sampling procedure detailed in \cref{sec:sampling} is applied to the approximation of $\oP_{t|t}$ and yields an \gls{opm} $\oP^{\s}_{t|t}$ characterised by the collection $\calL_t = \{(W^i_t,\calX^i_t)\}_{i=1}^{m_t}$ with
\eqns{
\calX^i_t = \big\{ \big( w^{i,j}_t, x^{i,j}_t \big) \big\}_{i=1}^{M^i_t}.
}
The algorithm can then be iterated by applying the prediction step described in \cref{ssec:prediction}.


The loss of diversity in the resampling step can be further reduced as follows: 
\begin{enumerate}
\item Compute $p^{i,j}_t = \calP_{\d}(\{w^{i,j}_t\}_{j=1}^{M^i_t})$ for all $i \in \{1,\dots,m_t\}$
\item Apply the standard resampling to the samples $(i,j)$ verifying $w^{i,j}_t = p^{i,j}_t$ only (the weights indicated in red in \cref{fig:waterPouringDiscrete})
\end{enumerate}
This simple modification ensures that resampling is only applied to samples with low weight, hence slowing down the decrease in sample diversity without introducing additional parameters. This modified resampling is referred to as the \emph{selective} resampling in \cref{sec:results}, as opposed to the basic procedure that resamples \emph{all} samples.

\subsection{Pseudo-code}

The complete \gls{smc} implementation of the \gls{opm} recursion given in \cref{eq:filteringUncertainTransition} is summarised in \cref{alg:multiPossibility}, where sampling applied to an approximated \gls{opm} $\oP^{\s}$ is understood as follows: sampling $f$ from the distribution $P$ (as described in line~\ref{alg:sampling:fsimP} of \cref{alg:sampling}) is replaced by selecting an index $i$ according to the weights $\{W_i\}$.

\begin{algorithm}
\caption{\Gls{smc} implementation for \gls{opm}-based filtering}
\label{alg:multiPossibility}
\begin{algorithmic}[1]
\Function{$\calL_T = \text{\textnormal{Possibility filter}}(\oP_{0,0},N)$}{} 
\State \# \emph{Initialisation}
\State $\calL_0 = \text{Approximation}(\oP_{0|0},N)$
\For {$t=1,\dots,T$}
\State \# \emph{Prediction}
\For {$i=1,\dots,m_{t-1}$}
\For {$j=1,\dots,M^i_{t-1}$}
\State $\tilde{x}^{i,j}_t \sim \calP_{\c}( g_t(\cdot \given x^{i,j}_{t-1}))$
\State Compute $w^{i,j}_{t|t-1}$ according to \cref{eq:predw}
\EndFor
\State Compute $W^i_{t|t-1}$ according to \cref{eq:predW}
\EndFor
\State \# \emph{Update}:
\For {$\ell=1,\dots,L_t$}
\State $s_{\ell} \sim S_t(\cdot \given X_t)$
\For {$i=1,\dots,m_{t-1}$}
\For {$j=1,\dots,M^i_{t-1}$}
\State Compute $\tilde{w}^{i,j,\ell}_t$ according to \cref{eq:upw}
\EndFor
\State Compute $\widetilde{W}^{i,\ell}_t$ according to \cref{eq:upW}
\EndFor
\EndFor
\State Define $\oP^{\s}_{t|t}$ according to \cref{eq:upPs}
\State \# \emph{Resampling}
\State $\calL_t = \text{Approximation}(\oP^{\s}_{t|t},N)$
\EndFor
\EndFunction
\end{algorithmic}
\end{algorithm}

In the case where all the involved \glspl{opm} are based on a single possibility function, the filtering recursion reduces to \cref{eq:singlePosPrediction} and \cref{eq:singlePosUpdate}. \Cref{alg:singlePossibility} details the corresponding simplified version of \cref{alg:multiPossibility}.

\begin{algorithm}
\caption{\Gls{smc} implementation for a single possibility}
\label{alg:singlePossibility}
\begin{algorithmic}[1]
\Function{$\calX_t = \text{\textnormal{Single-possibility filter}}(f_{0|0},N)$}{}
\State \# \emph{Initialisation}:
\For {$i=1,\dots,N$}
\State $x^i_0 \sim \calP_{\c}(f_{0|0})$
\State $w^i_0 \propto f_{0|0}(x^i_0)$
\EndFor
\For {$t=1,\dots,T$}
\State \# \emph{Prediction}:
\For {$i=1,\dots,N$}
\State $\tilde{x}^i_t \sim \calP_{\c}( g_t(\cdot \given x^i_{t-1}))$
\State $w^i_{t|t-1} \propto \max_j w^j_{t-1} g_t(\tilde{x}^i_t \given x^j_{t-1})$
\EndFor
\State \# \emph{Update}:
\For {$i=1,\dots,N$}
\State $\tilde{w}^i_t \propto w^i_{t|t-1} s_t(y_t \given \tilde{x}^i_t)$
\EndFor
\State \# \emph{Resampling}:
\For {$i=1,\dots,N$}
\State $a_i \sim \calP_{\d}\big(\{\tilde{w}^i_t\}_{i=1}^N\big)$
\State $x^i_t = \tilde{x}^{a_i}_t$
\State $w^i_t \propto \tilde{w}^{a_i}_t$
\EndFor
\State $\calX_t = \{(w^i_t,x^i_t)\}_{i=1}^N$
\EndFor
\EndFunction
\end{algorithmic}
\end{algorithm}

\section{Simulation results}
\label{sec:results}

In this section, the performance of the different design choices in the proposed approach are assessed and compared with a standard particle filter in simulations. The MAP is considered as an estimate for both filters since the mean is not always meaningful, e.g.\ when the posterior distribution is multi-modal. The MAP of the particle filter is obtained via
\eqnl{eq:trueMAP}{
\hat{x}_t = \argmax_{1 \leq i \leq N} p(y_t \given x^i_t) \sum_{j=1}^N p(x^i_t \given x^j_{t-1}) w^j_{t-1}
}
where $\{(x^i_{t-1},w^i_{t-1})\}_{i=1}^N$ is the collection of weighted particles before prediction and $\{x^i_t\}_{i=1}^N$ are the particles after update at time $t$, and where the usual abuse of notations using $p$ to denote the probability density function of any given argument is used. The sample with highest weight is simply considered as the MAP for the proposed approach.

\begin{remark}
When the dimension is low, an approximate MAP for the particle filter could also be calculated using kernel density estimation (KDE), that is
\eqns{
\hat{x} = \argmax_x \bigg( \dfrac{1}{Nh} \sum_{i=1}^N K\Big(\dfrac{x-x_i}{h}\Big) \bigg),
}
for some kernel $K$, some bandwidth $h > 0$ and some collection $\{x_i\}_{i=1}^N$ of i.i.d.\ samples. The analogous operation for the with possibility functions satisfy
\eqnsa{
\hat{x} & = \argmax_{x} \bigg( \max_{1 \leq i \leq N} w_i K'\Big(\dfrac{x-x_i}{h}\Big) \bigg) \\
& = x_j \quad\text{ with }\quad j = \argmax_{1 \leq i \leq N} w_i,
}
for some possibility function $K'$ reaching its maximum at $0$ and some collection of weighted samples $\{(w_i,x_i)\}_{i=1}^N$. This confirms that the sample with highest weight can be considered as the MAP when performing inference with possibility functions.
\end{remark}

\begin{table*}
\caption{Total \acrshort{rmse} and execution time (between brackets) for filtering based on probability (\emph{Pr}) and possibility (\emph{Po}) \emph{modelling}, with different sampling for \emph{continuous} and \emph{discrete} distributions (\emph{scaled}, \emph{global} entropy or \emph{local} entropy), different \emph{complexities} (linear or quadratic) and different \emph{resampling} techniques (based on \emph{all} samples or on \emph{selective} resampling). The results are averaged over $1000$ Monte Carlo runs. The two best overall filters are indicated in \textcolor{Orange}{orange} and the two best linear filters in \textcolor{RoyalBlue}{blue}, with the best in each category being in bold.}
\label{tbl:results}
\centering
{\footnotesize
\setlength{\tabcolsep}{.45em}
\begin{tabular}{ccccc|ccc|ccc}
& & & & & \multicolumn{3}{c|}{Scenario 1} & \multicolumn{3}{c}{Scenario 2} \\
\hline
Model & Continuous & Discrete & Complexity & Resampling & $N = 256$ & $N = 512$ & $N=1024$ & $N = 128$ & $N=256$ & $N = 512$ \\
\hline
Pr &   -    &   -    & $\calO(N)$   & All      & $42.34$ ($0.14$) & $42.83$ ($0.29$) & $43.04$ ($0.61$) & $29.61$ ($0.05$) & $29.76$ ($0.09$) & $30.08$ ($0.19$) \\
Po & Scaled & Scaled & $\calO(N)$   & Select.\ & $43.70$ ($0.25$) & $40.79$ ($0.50$) & $38.48$ ($1.04$) & $27.56$ ($0.10$) & \bestL{25.38} ($0.19$) & \bestL{24.20} ($0.38$) \\
Po & Global & Global & $\calO(N)$   & Select.\ & $56.67$ ($0.28$) & $54.05$ ($0.58$) & $51.50$ ($1.30$) & $32.29$ ($0.10$) & $29.30$ ($0.21$) & $28.11$ ($0.45$) \\
Po & Global & Global & $\calO(N)$   & All      & $56.84$ ($0.28$) & $54.17$ ($0.58$) & $51.06$ ($1.29$) & $32.11$ ($0.10$) & $29.26$ ($0.21$) & $28.39$ ($0.45$) \\
Po & Global & Local  & $\calO(N)$   & Select.\ & \bestL{40.11} ($0.25$) & \bestL{38.92} ($0.49$) & \bestL{38.10} ($1.00$) & \bestL{26.54} ($0.09$) & \sbestL{25.95} ($0.17$) & \sbestL{25.68} ($0.35$) \\
Po & Global & Local  & $\calO(N)$   & All      & \sbestL{40.58} ($0.25$) & \sbestL{39.07} ($0.50$) & \sbestL{38.17} ($1.02$) & \sbestL{26.66} ($0.09$) & $26.23$ ($0.18$) & $25.88$ ($0.37$) \\
\hline
Pr &   -    &   -    & $\calO(N^2)$ & All      & \best{38.35} ($0.65$) & \sbest{38.39} ($1.77$) & $38.64$ ($8.47$) & $26.39$ ($0.19$) & $26.60$ ($0.58$) & $26.79$ ($1.65$) \\
Po & Scaled & Scaled & $\calO(N^2)$ & Select.\ & $44.56$ ($0.64$) & $41.89$ ($1.75$) & $39.94$ ($8.46$) & $27.66$ ($0.19$) & $25.16$ ($0.57$) & $24.17$ ($1.63$) \\
Po & Global & Global & $\calO(N^2)$ & Select.\ & $56.83$ ($0.66$) & $53.78$ ($1.83$) & $50.54$ ($8.70$) & $28.68$ ($0.19$) & $25.32$ ($0.59$) & $23.64$ ($1.70$) \\
Po & Global & Global & $\calO(N^2)$ & All      & $56.33$ ($0.66$) & $53.45$ ($1.82$) & $50.51$ ($8.69$) & $28.54$ ($0.19$) & $25.41$ ($0.59$) & $23.69$ ($1.70$) \\
Po & Global & Local  & $\calO(N^2)$ & Select.\ & \sbest{39.86} ($0.63$) & \best{38.20} ($1.73$) & \best{37.06} ($8.33$) & \best{24.06} ($0.18$) & \best{22.67} ($0.55$) & \best{21.93} ($1.58$) \\
Po & Global & Local  & $\calO(N^2)$ & All      & $40.18$ ($0.63$) & $38.48$ ($1.74$) & \sbest{37.22} ($8.36$) & \sbest{24.91} ($0.18$) & \sbest{23.28} ($0.56$) & \sbest{22.34} ($1.61$)
\end{tabular}
}
\end{table*}

\subsection{Simulations in the single-function case}

In this section, we consider the case where all the involved \glspl{opm} are based on a single possibility function. The proposed method is, in this case, referred to as the \emph{(particle) possibility filter.}

\subsubsection{Scenario with Gaussian distributed noises}
\label{sssec:exactModel}

A standard 4-dimensional \gls{hmm} is first considered with linear-Gaussian dynamics (nearly constant velocity model in the plane with standard deviation $\sigma = 1\si{m/s^2}$) and linear-Gaussian observation (noisy observation of the position with standard deviation $\varsigma = 0.1\si{m}$) on a scenario with $100$ time steps of $\Delta = 0.1\si{s}$. The filtering equations then take the form
\eqnsa{
X_t & =
\begin{bmatrix}
\bsF & \zero_{2,2} \\
\zero_{2,2} & \bsF
\end{bmatrix}
X_{t-1} + \sigma U_t \\
Y_t & =
\begin{bmatrix}
1 & 0 & 0 & 0 \\
0 & 0 & 1 & 0
\end{bmatrix}
X_t + \varsigma U'_t
}
where $U_t \sim \calN(\zero_{4,1}, \bsQ)$ and $U'_t \sim \calN(\zero_{4,1}, \bsI_2)$ with $\zero_{d,d'}$ the zero matrix of dimension $d \times d'$ and $\bsI_{d}$ is the identity matrix of dimension $d\times d$, where
\eqns{
\bsF =
\begin{bmatrix}
1 & \Delta \\
0 & 1
\end{bmatrix}
\AND
\bsQ =
\begin{bmatrix}
\bsQ' & \zero_{2,2} \\
\zero_{2,2} & \bsQ'
\end{bmatrix}.
}
with
\eqns{
\bsQ' =
\begin{bmatrix}
\Delta^4/3 & \Delta^3/2 \\
\Delta^3/2 & \Delta^2
\end{bmatrix}.
}
The initial state $X_0$ is set to $[0, 1, 0, 1]^T$, and the initial variance is $0.01 \bsI_{4}$.

Both the particle and possibility filters are given these parameters, but the particle filter takes into account the fact that the underlying observation and dynamic noises are Gaussian whereas the possibility filter uses these parameters in a Gaussian possibility function of the form \cref{eq:GaussianPossibility}, which is less informative.

The results are shown in terms of \gls{rmse} in the column \emph{Scenario~1} of \cref{tbl:results} for different number of samples ($N=256, 512, 1024$). Out of all the different implementations of the possibility filter, the one using the global entropy method for continuous spaces, the local entropy approach for discrete space and the quadratic evaluation of the predicted weights performs the best for all numbers of samples. This implementation also shows a competitive performance when compared to the particle filter with true \gls{map} (indicated by a complexity of $\calO(N^2)$). This is a good result for the possibility filter since the information it takes is weaker than the one of the particle filter, in the sense that the possibility filter does not assume that the model is given exactly. The results are similar for linear implementations, with the global/local one performing the best overall. The \gls{rmse} obtained with the particle filter in the case where the particle with highest weight is considered as an estimate is also given as an indication (indicated by a complexity of $\calO(N)$). 



\subsubsection{Scenario with Student's t distributed noises}
\label{sssec:inexactModel}

In order to assess the performance of the proposed method in the presence of modelling discrepancies, a 2-dimensional \gls{hmm} is considered with linear dynamics and observation models that are similarly to the ones of the first scenario, i.e.\
\eqnsa{
X_t & =
\bsF 
X_{t-1} + \frac{\sigma}{\hat\sigma} U_t \\
Y_t & =
\begin{bmatrix}
1 & 0
\end{bmatrix}
X_t + \frac{\varsigma}{\hat\varsigma}U'_t
}
but where the noises are Student's~t distributed, i.e.\ $U_t \sim \calS_{\nu}$ and $U'_t \sim \calS_{\nu'}$ with $\calS_{\nu}$ the Student's~t distribution with $\nu$ degrees of freedom. The coefficients $\hat\sigma$ and $\hat\varsigma$ are defined as
\eqns{
\hat\sigma^2 = \dfrac{\nu}{\nu-2} \AND \hat\varsigma^2 = \dfrac{\nu'}{\nu'-2}
}
and help ensuring that the variance in the noise terms is the same as in the first scenario. We consider the values $\nu = \nu' = 5$ in the simulations. The initial state $X_0$ is set to $[0, 1]^T$, and the initial variance is $0.01 \bsI_{2}$.

The model communicated to the considered methods is however linear-Gaussian, with the means and variances of the actual noises. In this case, the two best implementations of the possibility filter show better performance than the particle filter, as shown in the \emph{Scenario 2} column of \cref{tbl:results}. As far as the linear filters are concerned, the implementation which is based on sampling from scaled possibility functions performs well with $256$ and $512$ samples. However, the global/local implementation remains competitive throughout and is therefore preferred. 




\subsection{Simulations in the general case}

In order to demonstrate the performance of the general \gls{smc} algorithm introduced in \cref{sec:SMC}, a case where the initial knowledge can be represented by multiple possibility functions is considered. In this section, only the best-performing implementation of the proposed method is evaluated.

The estimation of the angular position $\theta_t$ and rotation speed $\dot{\theta}_t$ of a fix point on a spinning disk is considered. The state space is $\boX = (-\pi,\pi] \times \bbR$ and the state is $x_t = [\theta_t, \dot{\theta}_t]^T$. The dynamics is modelled by a linear-Gaussian nearly-constant rotation speed model with a standard deviation of $1\si{rad/s^2}$. The considered observation model is
\eqns{
Y_t = \cos(\theta_t) + U'_t,
}
where $\{U'_t\}_{t \in \bbT}$ is a collection of normally-distributed random variables with a standard deviation of $0.1$. This observation implies that the posterior will be bi-modal if the prior information is not providing the direction of rotation. We therefore consider a prior knowledge of the form
\eqns{
\oP_0(\tf) = \dfrac{1}{2} \ninf{ \tf \cdot \bar\calN([0,1]^T, \bsP)} + \dfrac{1}{2} \ninf{ \tf \cdot \bar\calN([0,-1]^T, \bsP)}
}
with $\bsP$ the diagonal matrix corresponding to a standard deviation of $0.1\si{rad}$ in angular position and $0.2\si{rad/s}$ in rotation speed. The \gls{opm} $\oP_0$ models that the rotation is clockwise with a probability of $0.5$ and anti-clockwise otherwise. The corresponding prior for the particle filter is
\eqns{
p(x_0) = \dfrac{1}{2}\calN(x_0; [0,1]^T, \bsP) + \dfrac{1}{2}\calN(x_0; [0,-1]^T, \bsP).
}

\begin{figure}
\centering
\includegraphics[width=\columnwidth,trim = 105pt 250pt 110pt 270pt, clip]{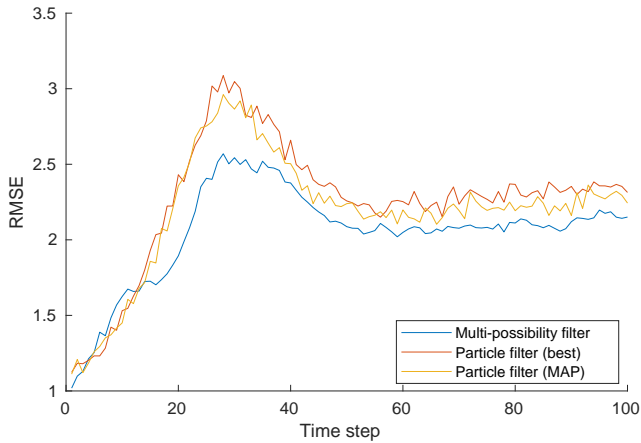}
\caption{\Gls{rmse} for the multi-possibility filter and particle filter on the spinning-disk scenario ($250$ samples, $1000$ Monte Carlo runs).}
\label{fig:MultiPos_Mod3}
\end{figure}

The results are shown in \cref{fig:MultiPos_Mod3} and indicate that the multi-possibility filter has a better performance than the particle filter at almost all iterations despite the fact that the exact model is given.

The scenario considered in this section is simple when compared to the capabilities of the multi-possibility filter in the sense that the number of possibility functions can be large and highly varying in general whereas this number is limited to $2$ in this example. However, the obtained results show that considering multiple possibilities can be beneficial even in this case.

\section{Conclusion}

Sequential Monte Carlo algorithms based on outer-measure recursions have been introduced and assessed in simulations. In particular, it has been shown that sample-based approximations can be used to propagate the \glspl{opm} corresponding to a stochastic filtering problem. As opposed to standard SMC methods, the samples are seen as support points on which the possibility functions underlying the considered \glspl{opm} are approximated. The benefits and the flexibility offered by the proposed method have been demonstrated in simulation in the presence of modelling discrepancies. Such a flexibility could be crucial in practice since the true model is rarely known for real data. Future work will consider how parameter estimation can be performed in the proposed framework.

\bibliographystyle{abbrv}
\bibliography{Methodology}

\appendix

\subsection{Proofs}

\begin{proof}[Proof of \cref{thm:predition}]
The \gls{opm} $\oP$ on $\boX_{t-1} \times \boX_t$ joining the \gls{opm} $\oP_{t-1|t-1}$ and the conditional \gls{opm} $\oQ_t(\cdot \given X_{t-1} = x')$ is characterised by
\eqnsml{
\oP(\bstf) = \int \sup_{(x',x) \in \boX_{t-1} \times \boX_t} \big( \bstf(x',x) g(x \given x') f(x') \big) \\
\times Q_t(\d g \given X_{t-1})  P_{t-1|t-1}(\d f)
}
for any $\bstf \in \boL^{\infty}(\boX_{t-1} \times \boX_t)$. The expression of the predicted \gls{opm} $\oP_{t|t-1}$ is deduced from the following marginalisation over $\boX_{t-1}$:
\eqnsa{
\oP_{t|t-1}(\tf) & = \oP(\ind{\boX_{t-1}} \times \tf) \\
& = \int \sup_{x \in \boX_t} \Big(\tf(x) \sup_{x' \in \boX_{t-1}} \big( g(x \given x') f(x') \big) \Big) \\
& \qquad\qquad\qquad\qquad \times Q_t(\d g \given X_{t-1})  P_{t-1|t-1}(\d f),
}
for any $\tf \in \boL^{\infty}(\boX_t)$, in which the mapping
\eqns{
\zeta_t(f,g)(x) = \sup_{x' \in \boX_{t-1}} g(x \given x') f(x')
}
can be identified, concluding the proof of the \lcnamecref{thm:predition}.
\end{proof}

\begin{proof}[Proof of \cref{thm:generalUpdate}]
The \gls{opm} $\oP$ on $\boX_t \times \boY_t$ joining the \gls{opm} $\oP_{t|t-1}$, the likelihood $\oS_t(\cdot \given X_t = x)$ and the observed information $\oI_t$ is characterised by
\eqnsml{
\oP(\bstf) = \int \sup_{(x,y) \in \boX_t \times \boY_t} \big( \bstf(x,y) h(y) s(y\given x) f(x) \big) \\
\times I_t(\d h) S_t(\d s \given X_t)  P_{t|t-1}(\d f)
}
for any $\bstf \in \boL^{\infty}(\boX_t \times \boY_t)$. Bayes formula  can be written for \glspl{opm} as
\eqns{
\oP_{t|t}(\tf) = \dfrac{\oP(\tf \times \ind{\boY_t})}{\oP(\ind{\boX_t \times \boY_t})},
}
for any $\tf \in \boL^{\infty}(\boX_t)$. The desired results follows from rewriting $\oP(\tf \times \ind{\boY_t})$ as
\eqnsml{
\oP(\tf \times \ind{\boY_t}) = \int \sup_{x \in \boX_t} \big( \tf(x) f(x) \ninf{ h \cdot s(\cdot \given x)} \big) \\
\times I_t(\d h) S_t(\d s \given X_t)  P_{t|t-1}(\d f)
}
and from noticing that $\oP(\ind{\boX_t \times \boY_t}) = \oP(\tf \times \ind{\boY_t}) |_{\tf = \ind{\boX_t}}$.
\end{proof}

\begin{proof}[Proof of \cref{res:convergence}]
First assume that $\ninf{ \tf \cdot f } = 0$, then either $w_i = 0$ or $\tf(x_i) = 0$ for all $i \in \{1,\dots,N\}$, in which case the result is obvious. Now assuming that $\ninf{ \tf \cdot f } > 0$, let $B_{\epsilon}$ be a subset of the $\epsilon$-neighbourhood of $A = \argmax_x \tf(x) f(x)$ for some $\epsilon > 0$. Since $\tf\cdot f$ is Lipschitz, there exists $K \geq 0$ such that $|w_i \tf(x_i) - \ninf{ \tf \cdot f } | \leq K |x_i -x^*|$ where $x^*$ is the element of $A$ that is closest to $x_i$, so that $x_i \in B_{\epsilon}$ implies $|w_i \tf(x_i) - \ninf{ \tf \cdot f } | \leq K\epsilon$.  It follows from the fact that the support of $p$ is equal to the one of $f$ that $p(B_{\epsilon}) > 0$. Therefore, it holds as required that for any $\epsilon > 0$ and any $\delta > 0$, there exists an integer $N'$ such that the probability for all the samples to be outside $B_{\epsilon}$ is smaller than $\delta$ for any $N \geq N'$.
\end{proof}

\begin{proof}[Proof of \cref{res:monotonePossibility}]
The ``if'' part of the statement follows from the definition of boundedness with the subset $(-\infty,x]$ since
\eqns{
F(x) = \int_{-\infty}^x p(y) \d y \leq \sup_{y \in (-\infty,x]} f(y) = f(x),
}
for any $x \leq b$. For the ``and only if'' part, it is sufficient to notice that for any $B \subseteq I$ if we let $x = \sup B$ then
\eqns{
\int_B p(y) \d y \leq F(x) \leq f(x) = \sup_{y \in B} f(y).
}
so that $\int_B p(y) \d y \leq \sup_{y \in B} f(y)$ as required.
\end{proof}

\begin{proof}[Sketch of proof for \cref{res:maxEntropy}]
If there is no constraint, the maximum-entropy probability distribution on an interval $[-a,0]$ of $\bbR$ for some $a > 0$ is the uniform distribution $\calU([-a,0])$. The corresponding \gls{cdf} is the affine function with value $0$ at $-a$ and $1$ at $0$. By symmetry, we simplify the problem to finding a function $p^*$ on $(-\infty,0]$ such that $p^* = \argmax_p H(p)$ subject to
\eqns{
\int_{-\infty}^x p(y) \d y = F(x) \leq \frac{1}{2}f(x)
}
for any $x \leq 0$, so that $p^*$ integrates to $1/2$ on $(-\infty,0]$. Indeed, we attribute half of the probability mass on the interval $(-\infty,0]$ by symmetry, and $f$ is monotonically increasing on this interval so that \cref{res:monotonePossibility} can be applied. This aspect is illustrated in \cref{fig:cdfOfPStar}. Although there is no uniform distribution on $(-\infty,0]$, the maximum-entropy distribution bounded by $f/2$ on this interval has a \gls{cdf} that is equal to $f/2$ on the interval $(-\infty,x^*]$ and is then affine on the interval $[x^*,0]$, where $x^*$ is the point at which the tangent to $f/2$ goes through the point $(0,1/2)$.
\end{proof}



\end{document}